\documentclass[10pt,aps,prl,amsmath,amssymb,twocolumn,%
				letterpaper,showpacs,footinbib,%
				balancelastpage,raggedbottom,
				citeautoscript,reprint,longbibliography,%
				floatfix]{revtex4-1}

\usepackage{graphicx}
\usepackage{dcolumn}
\usepackage{bm}

\usepackage{multirow}

\begin{document}

\title{First-principles study of the multi-mode anti-ferroelectric
  transition of PbZrO$_{3}$}

\author{Jorge \'I\~niguez,$^{1}$ Massimiliano Stengel,$^{2,1}$ Sergey
  Prosandeev,$^{3}$ and L. Bellaiche$^{3}$}

\affiliation{$^{1}$Institut de Ci\`encia de Materials de Barcelona
  (ICMAB-CSIC), Campus UAB, 08193 Bellaterra,
  Spain\\$^{2}$ICREA-Instituci\'o Catalana de Recerca i Estudis
  Avan\c{c}ats, 08010 Barcelona, Spain\\$^{3}$Physics Department and
  Institute for Nanoscience and Engineering, University of Arkansas,
  Fayetteville, Arkansas 72701, USA}

\begin{abstract}
We have studied {\sl ab initio} the phase transition in PbZrO$_{3}$, a
perovskite oxide usually presented as the prototypic
anti-ferroelectric material. Our work reveals the crucial role that
anti-ferrodistortive modes -- involving concerted rotations of the
oxygen octahedra in the structure -- play in the transformation, as
they {\em select} the observed anti-ferroelectric phase, among
competing structural variants, {\sl via} a cooperative trilinear
coupling.
\end{abstract}

\pacs{77.80.-e, 77.84.-s, 63.20.dk, 71.15.Mb}






\maketitle

From a structural point of view, most {\sl AB}O$_{3}$ perovskite
oxides present phases that can be regarded as distorted versions of
the cubic prototype \cite{glazer72,glazer75,giaquinta94}. Many are
characterized by concerted rotations of the O$_{6}$ octahedra that
constitute the basic building block of the lattice (e.g., SrTiO$_{3}$,
most manganites and nickelates). A second group presents off-centering
displacements of the {\sl A} and {\sl B} cations, which typically
result in a switchable ferroelectric (FE) polarization (e.g.,
BaTiO$_{3}$ or PbTiO$_{3}$). Finally, these two features appear
combined in some cases (e.g., BiFeO$_{3}$ or LiNbO$_{3}$). Exceptions
to these typical situations are uncommon, but are attracting growing
interest. In particular, materials displaying an anti-polar cation
displacement pattern, i.e. the so-called {\em anti-ferroelectric}
(AFE) order, are currently a focus of attention for both fundamental
and applied reasons \cite{rabe-inbook13,hao13,zhu12}.

Many phases characterized by rotations of the O$_{6}$ octahedra
(anti-ferrodistortive or AFD modes henceforth) also display anti-polar
displacements of the {\sl A} cations. Such anti-polar distortions are
typically a consequence of the AFD modes \cite{bellaiche13} and would
not exist in their absence; the ensuing AFE order can thus be regarded
as {\em improper} in nature. In contrast, here we are interested in
materials usually described as {\em proper} AFEs, in which a phase
transition accompanied by a dielectric anomaly is supposedly driven by
a primary anti-polar order parameter. (We adopt the most common
definition of a proper AFE transition~\cite{rabe-inbook13}.)
PbZrO$_{3}$ (PZO) displays such a striking behavior, and is usually
presented as the prototypic AFE crystal
\cite{sawaguchi51,shirane51,haun89,liu11}. However, the nature of
PZO's transition is far from being settled, as new experimental
results and conflicting physical pictures have recently been reported
\cite{tagantsev13,hlinka14}. Hence, there is a need to clarify PZO's
behavior and status as a model AFE. Here we present a first-principles
investigation to that end.

\begin{figure}
\includegraphics[width=\columnwidth]{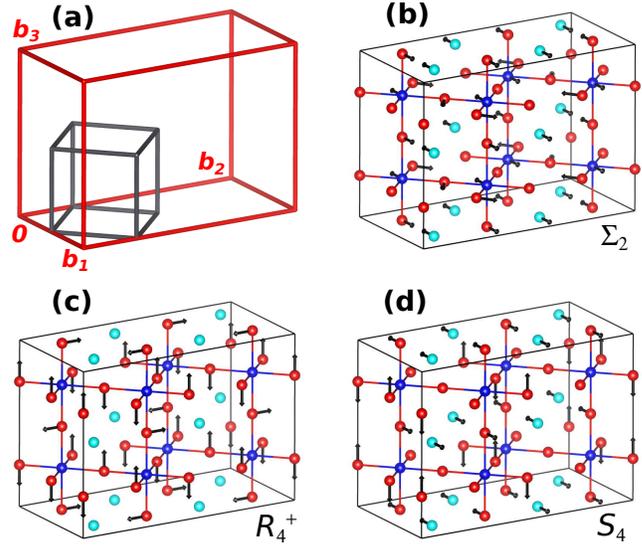}
\caption{(Color online) (a): Sketch of the 5-atom primitive cubic cell
  (small cube) and the $\sqrt{2}\times 2\sqrt{2}\times 2$ cell
  corresponding to PZO's AFE phase (large polyhedron). For the latter,
  the ${\bf b}_{i}$ lattice vectors are shown (see text). (b), (c),
  and (d): Three main distortion modes that lead to PZO's ground state
  (see text). Symmetry labels indicated. Bonds between Zr (dark blue)
  and O (red) atoms are drawn. Note the characteristic AFE
  displacement of the Pb (cyan) atoms associated with the $\Sigma_{2}$
  distortion.}
\label{fig1}
\end{figure}

{\sl Main modes and their couplings}.-- We followed the usual
first-principles approach to the investigation of a non-reconstructive
phase transition, taking advantage of the experimental knowledge of
the high-temperature (cubic $Pm\bar{3}m$, with the elemental 5-atom
unit cell) and low-temperature (orthorhombic $Pbam$, with a 40-atom
unit cell) structures \cite{teslic98,fujishita03}. The unit cell of
the AFE $Pbam$ phase can be viewed as a $\sqrt{2}\times
2\sqrt{2}\times 2$ multiple of the elemental unit, as sketched in
Fig.~\ref{fig1}(a). More precisely, if ${\bf a}_{1} = a(1,0,0)$, ${\bf
  a}_{2} = a(0,1,0)$, and ${\bf a}_{3} = a(0,0,1)$ define the ideal
5-atom cell in a Cartesian reference, the $Pbam$ cell vectors are
given by ${\bf b}_{1} = {\bf a}_{1}-{\bf a}_{2}$, ${\bf b}_{2} =
2({\bf a}_{1}+{\bf a}_{2})$, and ${\bf b}_{3} = 2{\bf a}_{3}$. As
shown in Fig.~\ref{fig1}(b), PZO's AFE distortion involves
displacements of the Pb cations along the pseudo-cubic $[1\bar{1}0]$
direction, modulated according to the wave vector ${\bf q} = {\bf
  q}_{\Sigma} = 2\pi (1/4,1/4,0)/a$ of the first Brillouin Zone (BZ)
of the 5-atom cell. (We give all vectors in the pseudo-cubic setting.)

We used standard methods based on density functional theory (details
in \cite{supp}) to relax the $Pm\bar{3}m$ and $Pbam$ phases, and got
results \cite{supp} in good agreement with experiments
\cite{teslic98,fujishita03} and previous theoretical works
\cite{singh95,piskunov07}. Then, we employed standard crystallographic
tools \cite{isotropy} to identify the symmetry-adapted distortions
connecting these two structures, and found three dominating ones
\cite{fnlabels}: (1) A $\Sigma_{2}$ component associated with the
${\bf q}_{\Sigma}$ wave vector and which captures 36.0~\% of the total
distortion. This is the AFE$_{\Sigma}$ displacement pattern sketched
in Fig.~\ref{fig1}(b). Interestingly, this distortion also has certain
AFD character, as oxygen displacements reminiscent of an in-phase
rotation about the ${\bf b}_{3}$ axis are clearly appreciated. (2) A
$R_{4}^{+}$ component associated to ${\bf q}_{R} = 2\pi
(1/2,1/2,1/2)/a$ and involving anti-phase rotations of the O$_{6}$
octahedra about the $[1\bar{1}0]$ axis; see sketch in
Fig.~\ref{fig1}(c). This AFD distortion amounts to 59.7~\% of the
total. (3) A $S_{4}$ distortion associated to ${\bf q}_{S} = 2\pi
(1/4,1/4,1/2)/a$ that captures 4.1~\% of the total and is sketched in
Fig.~\ref{fig1}(d). This distortion too has a mixed character,
combining AFE features with others that are reminiscent of AFD modes.

\begin{figure}
\includegraphics[width=0.9\columnwidth]{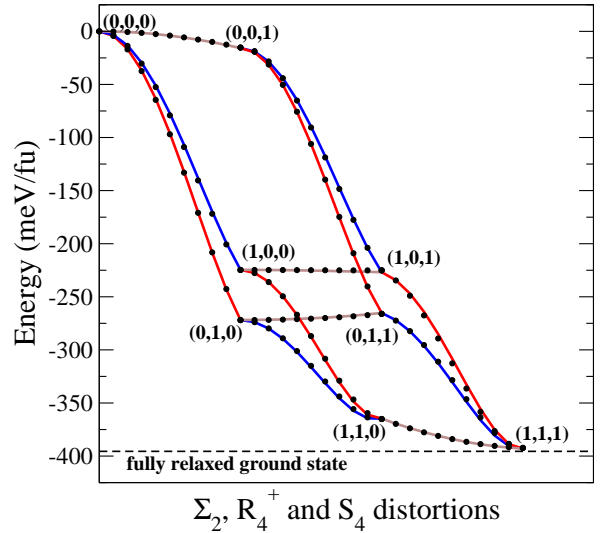}
\caption{(Color online) Circles: Computed energies of various
  structures defined by $(Q_{\Sigma},Q_{R},Q_{S})$ triads (see text;
  the triads for the limit and intermediate structures are
  indicated). Lines: Fit to Eq.~(\ref{eq:energy}); fitted parameters
  (meV {\sl per} formula unit): $A_{\Sigma}$~=~$-$337.3,
  $B_{\Sigma}$~=~112.7, $A_{R}$~=~$-$420.9, $B_{R}$~=~149.1,
  $A_{S}$~=~$-$16.4, $B_{S}$~=~0.6, $C_{\Sigma RS}$~=~$-$48.4,
  $D_{\Sigma R}$~=~131.7, $D_{\Sigma S}$~=~22.4, and
  $D_{RS}$~=~13.6. For all triads we considered the equilibrium cell
  of the $Pbam$ phase in the calculations (note that the $Q$'s do not
  involve any strain); results using the relaxed cubic cell are very
  similar \cite{supp}. The dashed horizontal line marks the energy of
  the fully relaxed ground state.}
\label{fig2}
\end{figure}

We then considered the simplest model that captures how the energy of
PZO changes as a function of these three distortions of the cubic
phase. Let $Q_{\Sigma}$, $Q_{R}$, and $Q_{S}$ denote the respective
amplitudes, which we normalize so that $Q_{\Sigma} = Q_{R} = Q_{S} =
1$ describe the situation at the $Pbam$ ground state. To fourth order,
the energy has the form \cite{isotropy}:
\begin{equation}
\begin{split}
E = & \; E_{\rm cubic} +\\
& A_{\Sigma} Q_{\Sigma}^{2} + B_{\Sigma} Q_{\Sigma}^{4} + A_{R}
Q_{R}^{2} + B_{R} Q_{R}^{4} + \\
& A_{S} Q_{S}^{2} + B_{S} Q_{S}^{4} + C_{\Sigma RS} Q_{\Sigma} Q_{R}
Q_{S} + \\
& D_{\Sigma R} Q_{\Sigma}^{2}Q_{R}^{2} + D_{\Sigma S}
Q_{\Sigma}^{2}Q_{S}^{2} + D_{RS} Q_{R}^{2}Q_{S}^{2} \, .
\end{split}
\label{eq:energy}
\end{equation}
Interestingly, we find a trilinear coupling $C_{\Sigma RS}$ involving
all three modes under consideration. Such a coupling, whose existence
had been noticed already \cite{hlinka14,prosandeev14}, is cooperative
in nature and provides a mechanism for the simultaneous occurrence of
the involved distortions. In order to get a quantitative estimate of
the model parameters, we investigated various paths to transit between
the $Q_{\Sigma} = Q_{R} = Q_{S} = 0$ and $Q_{\Sigma} = Q_{R} = Q_{S} =
1$ states. Figure~\ref{fig2} shows the results; the quality of the fit
to Eq.~(\ref{eq:energy}) (parameters in caption) is excellent
\cite{fn4thorder}.

Several conclusions can be drawn from these results. First, all three
$\Sigma_{2}$, $R_{4}^{+}$, and $S_{2}$ modes are instabilities of the
cubic phase when considered individually. The AFD instability is the
dominant and strongest one, closely followed by the AFE mode; in
contrast, the energy gain associated with the S$_4$ mode is clearly
smaller. Next, note that all the biquadratic couplings in
Eq.~(\ref{eq:energy}) are {\em competitive} in nature, i.e., we have
$D_{\Sigma R}, D_{\Sigma S}, D_{RS} > 0$. Interestingly, the AFE and
AFD instabilities are strong enough to coexist in spite of their
mutual repulsion; in contrast, condensation of either the AFE or AFD
distortion results in the stabilization of the $S_{4}$ mode. Finally,
and remarkably, the trilinear term has a considerable impact in the
ground state energy: when going from $(1,1,0)$ to $(1,1,1)$ via
relaxation of the $S_{4}$ distortion, the energy undergoes a
considerable reduction, by about 27~meV per formula unit (fu). These
three distortions together produce an energy gain of 392~meV/fu
compared to the cubic phase. Full relaxation (via other structural
modes not considered here) lead to a further energy decrease of only
3~meV/fu, which confirms the dominant character of $Q_{\Sigma}$,
$Q_{R}$ and $Q_{S}$.

\begin{table}[t!]
\caption{Computed energies of PZO adopting several structures
  characterized by the presence of AFD (labeled using Glazer's
  notation \cite{glazer72}) and polar distortions. For the polar
  distortions, we indicate whether they are FE or AFE and the
  (approximate in some cases) direction along which the dipoles
  lie. In cases in which there is more than one AFE component, we only
  indicate the dominant wave vector. The structures with asterisk have
  the same space group as the ground state (gs), but their
  corresponding cells are different.}

\vskip 1mm 

\begin{tabular*}{\columnwidth}{@{\extracolsep{\fill}}lccc}
\hline\hline
structure & \begin{tabular}{@{}c@{}}$E-E_{\rm gs}$\\(meV/fu)\end{tabular} 
& AFD dist. & polar dist. \\
\hline
$Pbam$ (gs) & 0   & $a^{-}a^{-}c^{0}$ & \begin{tabular}{@{}c@{}}
AFE $[1\bar{1}0]$\\ ${\bf q}=2\pi(1/4,1/4,0)/a$\end{tabular}\\
$Pm\bar{3}m$& 375 & --              & --                        \\
$I4/mcm$    & 113 & $a^{-}b^{0}b^{0}$ & --                        \\
$Imcm$      & 52  & $a^{-}a^{-}c^{0}$ & \begin{tabular}{@{}c@{}}
AFE $[1\bar{1}0]$\\ ${\bf q}=2\pi(1/2,1/2,1/2)/a$\end{tabular}\\
$R\bar{3}c$ & 75  & $a^{-}a^{-}a^{-}$ & --                        \\
$Pnma$      & 35  & $a^{-}a^{-}c^{+}$ & \begin{tabular}{@{}c@{}}
AFE $\sim [110]$\\ ${\bf q}=2\pi(0,0,1/2)/a$\end{tabular}\\
$P4mm$      & 124 & --              & FE $[100]$                \\
$Amm2$      & 91  & --              & FE $[110]$                \\
$R3m$       & 58  & --              & FE $[111]$                \\
$Ima2$      & 19  & $a^{-}a^{-}c^{0}$ & FE $[110]$                \\
$R3c$       & 7   & $a^{-}a^{-}a^{-}$ & FE $[111]$                \\
$Pbmm$      & 217 & --              & \begin{tabular}{@{}c@{}}
AFE $[1\bar{1}0]$\\ ${\bf q}= 2\pi(1/2,1/2,0)/a$\end{tabular}\\
$Pbam^{*}$   & 68  & --              & \begin{tabular}{@{}c@{}}
AFE $\sim [100]$\\ ${\bf q}=2\pi(1/4,1/4,0)/a$\end{tabular}\\
$Pbam^{*}$   & 60  & --              & \begin{tabular}{@{}c@{}}
AFE $\sim [100]$\\${\bf q}= 2\pi (1/8,1/8,0)/a$\end{tabular}\\
\hline\hline
\end{tabular*}
\label{tab1}
\end{table}

{\sl Competing phases}.-- Earlier studies suggest that the equilibrium
$Pbam$ phase of PZO competes with other FE, AFE, and AFD structures
\cite{shirane52,singh95,kagimura08,reyeslillo13,prosandeev14}. We thus
ran a series of computer experiments to shed some light on why PZO
chooses such an unusual phase over more common alternatives.

We computed the energy of several hypothetical PZO phases displaying
combinations of the most typical AFD and FE displacement patterns. We
also considered a number of AFE patterns similar to AFE$_{\Sigma}$ but
with different periodicities. To obtain the AFE$_{\bf q}$ states, we
first displaced the Pb cations by hand along the $[1\bar{1}0]$
direction with the spatial modulation of the targeted wave vector, and
then let the structure fully relax while preserving the initial
symmetry. (In some cases the relaxed structures present more complex
Pb displacements than initially expected.)  From the results, which
are summarized in Table~\ref{tab1}, a number of conclusions emerge. On
one hand, the pure FE state $R3m$ is more stable than the purely AFD
or AFE ones, although they all lie relatively far from the ground
state (e.g., the $R3m$ solution lies at 58~meV/fu above). Combinations
of FE and AFD distortions, on the other hand, can produce structures
that are remarkably close to the ground state (the $R3c$ and $Ima2$
solutions lie at 7~meV/fu and 19~meV/fu, respectively). In fact, the
energy difference between the $Pbam$ and $R3c$ or $Ima2$ phases is
smaller than the energy associated with the trilinear term discussed
above (27~meV/fu), which suggests that such a coupling is essential
for the stabilization of the ground state. The central role played by
the mutual interaction between modes is also corroborated by our
results for the AFE$_{\bf q}$ phases: ${\bf q}_{\Sigma}$ is not the
most favorable modulation, suggesting that, in absence of the
$R_{4}^{+}$ and $S_{4}$ modes, there would be no reason for the
crystal to favor it over, e.g., ${\bf q} = 2\pi(1/8,1/8,0)/a$.

{\sl Phonon mode analysis}.-- The results discussed so far point to a
crucial role played by the AFD distortions in stabilizing the observed
antiferroelectric phase of PZO. It is therefore reasonable to
speculate that condensation of the AFD modes may have an important
impact on the unstable phonon branch associated with the polar
(FE/AFE) modes. To verify this hypothesis, we calculated the phonon
spectrum of PZO in two different configurations: first in the
optimized cubic phase, and second in a phase where we optimized (by
freezing it in by hand with an amplitude corresponding to
$Q_{R}$=1.13) the relevant AFD ($a^{-}a^{-}c^{0}$) component, while
keeping the other structural parameters identical to those of the
cubic geometry.

\begin{figure}
\includegraphics[width=\columnwidth]{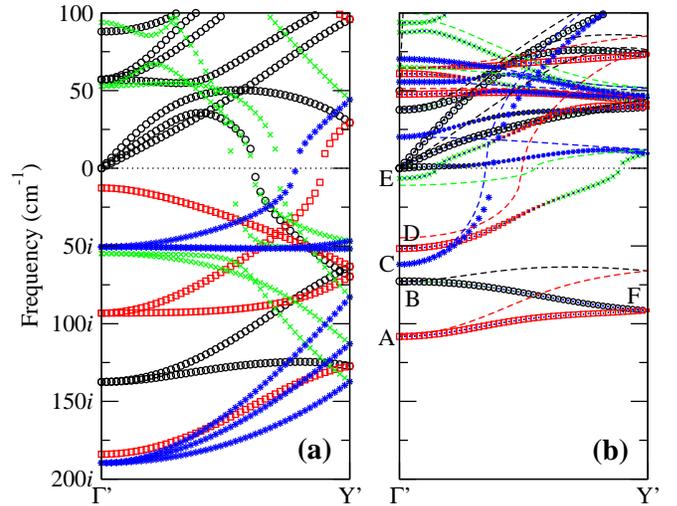}
\caption{(Color online) Phonon bands calculated along the
  $\Gamma'$-$Y'$ line in the BZ of the 20-atom ($\sqrt{2} \times
  \sqrt{2} \times 2$) cell used to simulate the $Q_{R}$-distorted
  structure (see text). Primed labels denote wave vectors of this BZ.
  As a consequence of the folding, such a path encompasses four
  inequivalent segments in the BZ of the primitive cubic cell, each of
  which has been assigned a different color and symbol type:
  $\Gamma$-$\Sigma$ (black circles), $M$-$\bar{\Sigma}$ (red squares),
  $Z$-$S$ (green crosses), and $R$-$\bar{S}$ (blue stars), where ${\bf
    q}_{\bar{\Sigma}} = - {\bf q}_{\Sigma}$ and ${\bf q}_{\bar{S}} =
  -{\bf q}_{S}$. The size of each symbol corresponds to the
  decomposition (projection) of a given mode into the four components
  described above. (a): Cubic phase ($Q_{R}=0$). (b): Distorted phase
  ($Q_{R}=1.13$). The dashed lines in (b) indicate the hypothetical
  phonon spectrum obtained by suppressing the interaction
  (off-diagonal terms) between ${\bf q}$-point pairs (see text). Only
  the low-energy part of the spectrum, including the unstable
  branches, is shown.}
\label{fig3}
\end{figure}

The resulting phonon bands, calculated on a $\Gamma'$-$Y'$ path in
reciprocal space, are shown in Fig.~\ref{fig3}. At $Q_{R}=0$
[Fig.~\ref{fig3}(a)] we recover the spectrum of the cubic reference,
appropriately folded onto the smaller BZ of the $\sqrt{2} \times
\sqrt{2} \times 2$ cell that we used to accomodate the AFD tilts. As
previous authors noted \cite{ghosez99}, this phase presents multiple
strong instabilities of both polar and AFD character, with the latter
(particularly those at ${\bf q}_{R}$) appearing to dominate over the
former. As expected, we observe a whole band of unstable polar
distortions, which includes the FE and AFE patterns associated with
most of the phases of Table~\ref{tab1}. Within this band, the FE mode
at $\Gamma'$ clearly dominates over the AFE modes.

The onset of the AFD distortions has a dramatic influence on the
unstable region of the phonon spectrum [Figure~\ref{fig3}(b)]. Most of
the formerly unstable modes have been stabilized -- only five branches
with imaginary frequency persist. At the $\Gamma'$ point of the folded
BZ these are (from most to least unstable): (A) a $c^{+}$-type AFD
mode that would lead to the $Pnma$ structure of Table~\ref{tab1}; (B)
a FE mode with the polarization oriented along $[1\bar{1}0]$ that
would lead to the $Ima2$ phase; (C) a $c^{-}$-type AFD mode leading to
$R\bar{3}c$; (D) an AFE mode with ${\bf q}= 2\pi(1/2,1/2,0)/a$ and
polar distortions aligned with $[1\bar{1}0]$; (E) a similar AFE mode
but with ${\bf q}= 2/\pi(0,0,1/2)/a$ that is also present in the
$Pnma$ phase. All of these branches, except two, eventually become
stable when moving away from $\Gamma'$. The surviving doublet (F) is
predominantly of AFE$_{\Sigma}$ character (80\%) with a smaller
$S_{4}$ component (20\%). This is, of course, the distortion that
directly leads to the ground-state structure. Remarkably, the doublet
is now {\em more unstable} than the FE mode at $\Gamma'$ [only
  AFD-$c^{+}$ appears to be stronger in panel~(b)], confirming the
crucial role played by $Q_{\rm R}$ in {\em selecting} a specific AFE
ground state \cite{fnPnma}.

To understand which couplings are primarily responsible for such an
outcome, we performed a further computational experiment. Note that,
to leading order, the effect of a certain $Q_{R} = \bar{Q}_{R}$
distortion on the phonon spectrum is given by two types of terms:
trilinear ones of the form $Q_{{\bf q}_{1},s}Q_{{\bf
    q}_{2},s'}\bar{Q}_{R}$, where $s$ and $s'$ label atomic
displacements and ${\bf q}_{1}+{\bf q}_{2}+{\bf q}_{R}$ is a
reciprocal lattice vector; biquadratic ones of the form $Q_{{\bf
    q},s}Q_{{\bf q},s'}^{*}\bar{Q}_{R}^{2}$. Hence, we recalculated
the phonon bands of the $\bar{Q}_{R}$-distorted structure while
suppressing the off-diagonal elements of the dynamical matrix that
couple $({\bf q}_{1},{\bf q}_{2})$ pairs. The result, shown as dashed
lines in Fig.~\ref{fig3}(b), clearly shows that the trilinear terms
play a crucial role in pushing the zone-boundary doublet (F) to lower
energies than the FE mode at $\Gamma'$, and effectively favors the
occurrence of the former in the ground state.

{\sl Nature of PZO's phase transition}.-- PZO's AFE structure involves
distortions of various symmetries, and it seems experimentally proved
that all of them appear spontaneously at a single,
strongly-discontinuous transformation upon cooling from the cubic
phase \cite{liu11,rabe-inbook13}. Our simulations do not incorporate
thermal effects, which prevents us from discussing PZO's transition in
a conclusive way. Yet, they provide some hints that are useful in
light of recent experimental measurements in which simultaneous
softening of multiple phonon modes was observed
\cite{tagantsev13,hlinka14}. Based on the fact that the acoustic
branch was also involved in the softening, theoretical models have
been formulated where the phase transition is rationalized in terms of
flexoelectric couplings \cite{tagantsev13}. Note that in this recent
model the ferroelectric branch is believed to drive the transition
and, in line with earlier works \cite{haun89}, the AFD distortions are
assumed to play a secondary role.

Let us begin by noting that, in principle, a strongly first-order
transition like PZO's may or may not be accompanied by a significant
mode softening. Our results are compatible with such a {\em
  quasi-reconstructive} transformation, of the sort that occurs in
perovskite oxides sharing some of PZO peculiar features. A notable
example is BiFeO$_{3}$
\cite{catalan09,dieguez11,prosandeev13,cazorla13}, which also presents
stable phases lying far below the cubic phase, strong FE, AFE, and AFD
instabilities, low-lying $R3c$ and $Pnma$ polymorphs, etc.

Notwithstanding these considerations, a particular soft mode might
still drive the transformation. Let us assume that the AFD $R_{4}^{+}$
mode plays such a role. (Our argument remains valid if we swap the
roles of the $R_{4}^{+}$ and $\Sigma_{2}$ modes.) Then, let us
consider the harmonic part of a Landau potential for the $\Sigma_{2}$
and $S_{4}$ modes, at the AFD transition temperature ($T_{R}$) at
which a $R_{4}^{+}$ distortion of amplitude $\bar{Q}_{R}$
materializes. We would have:
\begin{equation}
\begin{split}
F(Q_{\Sigma},Q_{S}) \sim \, & [ \tilde{A}_{\Sigma} \,
  (T_{R}-T_{\Sigma}) + \tilde{D}_{\Sigma R} \, \bar{Q}_{R}^{2}] \,
Q_{\Sigma}^{2} + \\
& [\tilde{A}_{S} \, (T_{R}-T_{S}) + \tilde{D}_{RS} \,
    \bar{Q}_{R}^{2}] \, Q_{S}^{2} + \\
& (\tilde{C}_{\Sigma RS} \bar{Q}_{R}) \, Q_{\Sigma} Q_{S} \, ,
\end{split}
\end{equation}
where $T_{\Sigma}$ and $T_{S}$ are the critical temperatures for the
corresponding instabilities, and all the $\tilde{A}$ and $\tilde{D}$
parameters are positive. Our calculations show that the AFD and
AFE$_{\Sigma}$ instabilities are similarly strong; thus, we can
tentatively assume $T_{R} \gtrsim T_{\Sigma}$. In such conditions, if
$[\tilde{A}_{\Sigma} \, (T_{R}-T_{\Sigma}) + \tilde{D}_{\Sigma R} \,
  \bar{Q}_{R}^{2}]$ is comparable in magnitude to $\tilde{C}_{\Sigma
  RS} \bar{Q}_{R}$, then the above harmonic term might present a
combined $\Sigma_{2} + S_{4}$ instability. In other words, the
occurrence of a discontinuous AFD transition could {\em spark} the
$\Sigma_{2}$ and $S_{4}$ distortions, resulting in a single
$Pm\bar{3}m \rightarrow Pbma$ transformation. (Alternatively, if the
conditions for the $\Sigma_{2} + S_{4}$ instability occurred at
$T_{\Sigma S} \lesssim T_{R}$, an intermediate phase could be observed
between $T_{\Sigma S}$ and $T_{R}$.)

Regardless of whether we should view PZO's transition as a
quasi-reconstructive one or a soft-mode driven one, the central role
of the AFD modes (to stabilize the $Pbam$ phase over other
alternatives or provide the induction mechanism through a trilinear
coupling, respectively) is clear from our results. This is in line
with the hypotheses made in \cite{hlinka14} and questions the validity
of models that treat the AFD distortions as a secondary effect
\cite{tagantsev13}.

In summary, our work has led to key insights concerning PZO's AFE
transition. Most importantly, our results suggest that the multi-mode
character of the transformation is essential to its very
occurrence. In particular, we have found that AFD modes associated to
oxygen octahedra rotations play a key role in {\em selecting} the
experimentally observed AFE structure, over competing structural
variants, via a cooperative trilinear coupling. Consequently, it does
not seem appropriate to think of a primary AFE order parameter driving
PZO's transition; rather, our results suggest a peculiar improper-like
nature of the AFE order. PZO thus appears as a very complex material,
which poses the provocative question of whether its intricate behavior
can be taken as representative of anti-ferroelectricity in perovskite
oxides.

Work supported by MINECO-Spain (Grants No. MAT2010-18113 and
No. CSD2007-00041). We also acknowledge support from Generalitat de
Catalunya (2014 SGR 301) (J\'I and MS) and ONR Grants
No. N00014-11-1-0384 and No. N00014-12-1-1034 (SP and LB). We used the
facilities provided by the CESGA supercomputing center. Some figures
were prepared using VESTA \cite{vesta}.



\begin{thebibliography}{41}%
\makeatletter
\providecommand \@ifxundefined [1]{%
 \@ifx{#1\undefined}
}%
\providecommand \@ifnum [1]{%
 \ifnum #1\expandafter \@firstoftwo
 \else \expandafter \@secondoftwo
 \fi
}%
\providecommand \@ifx [1]{%
 \ifx #1\expandafter \@firstoftwo
 \else \expandafter \@secondoftwo
 \fi
}%
\providecommand \natexlab [1]{#1}%
\providecommand \enquote  [1]{``#1''}%
\providecommand \bibnamefont  [1]{#1}%
\providecommand \bibfnamefont [1]{#1}%
\providecommand \citenamefont [1]{#1}%
\providecommand \href@noop [0]{\@secondoftwo}%
\providecommand \href [0]{\begingroup \@sanitize@url \@href}%
\providecommand \@href[1]{\@@startlink{#1}\@@href}%
\providecommand \@@href[1]{\endgroup#1\@@endlink}%
\providecommand \@sanitize@url [0]{\catcode `\\12\catcode `\$12\catcode
  `\&12\catcode `\#12\catcode `\^12\catcode `\_12\catcode `\%12\relax}%
\providecommand \@@startlink[1]{}%
\providecommand \@@endlink[0]{}%
\providecommand \url  [0]{\begingroup\@sanitize@url \@url }%
\providecommand \@url [1]{\endgroup\@href {#1}{\urlprefix }}%
\providecommand \urlprefix  [0]{URL }%
\providecommand \Eprint [0]{\href }%
\providecommand \doibase [0]{http://dx.doi.org/}%
\providecommand \selectlanguage [0]{\@gobble}%
\providecommand \bibinfo  [0]{\@secondoftwo}%
\providecommand \bibfield  [0]{\@secondoftwo}%
\providecommand \translation [1]{[#1]}%
\providecommand \BibitemOpen [0]{}%
\providecommand \bibitemStop [0]{}%
\providecommand \bibitemNoStop [0]{.\EOS\space}%
\providecommand \EOS [0]{\spacefactor3000\relax}%
\providecommand \BibitemShut  [1]{\csname bibitem#1\endcsname}%
\let\auto@bib@innerbib\@empty
\bibitem [{\citenamefont {Glazer}(1972)}]{glazer72}%
  \BibitemOpen
  \bibfield  {author} {\bibinfo {author} {\bibfnamefont {A.~M.}\ \bibnamefont
  {Glazer}},\ }\href@noop {} {\bibfield  {journal} {\bibinfo  {journal} {Acta
  Crystallographica B}\ }\textbf {\bibinfo {volume} {28}},\ \bibinfo {pages}
  {3384} (\bibinfo {year} {1972})}\BibitemShut {NoStop}%
\bibitem [{\citenamefont {Glazer}(1975)}]{glazer75}%
  \BibitemOpen
  \bibfield  {author} {\bibinfo {author} {\bibfnamefont {A.~M.}\ \bibnamefont
  {Glazer}},\ }\href@noop {} {\bibfield  {journal} {\bibinfo  {journal} {Acta
  Crystallographica A}\ }\textbf {\bibinfo {volume} {31}},\ \bibinfo {pages}
  {756} (\bibinfo {year} {1975})}\BibitemShut {NoStop}%
\bibitem [{\citenamefont {Giaquinta}\ and\ \citenamefont {zur
  Loye}(1994)}]{giaquinta94}%
  \BibitemOpen
  \bibfield  {author} {\bibinfo {author} {\bibfnamefont {D.~M.}\ \bibnamefont
  {Giaquinta}}\ and\ \bibinfo {author} {\bibfnamefont {H.-C.}\ \bibnamefont
  {zur Loye}},\ }\href@noop {} {\bibfield  {journal} {\bibinfo  {journal}
  {Chemistry of Materials}\ }\textbf {\bibinfo {volume} {6}},\ \bibinfo {pages}
  {365} (\bibinfo {year} {1994})}\BibitemShut {NoStop}%
\bibitem [{\citenamefont {Rabe}(2013)}]{rabe-inbook13}%
  \BibitemOpen
  \bibfield  {author} {\bibinfo {author} {\bibfnamefont {K.~M.}\ \bibnamefont
  {Rabe}},\ }\enquote {\bibinfo {title} {Antiferroelectricity in oxides: A
  reexamination},}\ in\ \href {\doibase 10.1002/9783527654864.ch7} {\emph
  {\bibinfo {booktitle} {Functional Metal Oxides}}}\ (\bibinfo  {publisher}
  {{Wiley-VCH Verlag GmbH \& Co. KGaA}},\ \bibinfo {year} {2013})\ pp.\
  \bibinfo {pages} {221--244}\BibitemShut {NoStop}%
\bibitem [{\citenamefont {Hao}(2013)}]{hao13}%
  \BibitemOpen
  \bibfield  {author} {\bibinfo {author} {\bibfnamefont {X.}~\bibnamefont
  {Hao}},\ }\href@noop {} {\bibfield  {journal} {\bibinfo  {journal} {Journal
  of Advanced Dielectrics}\ }\textbf {\bibinfo {volume} {3}},\ \bibinfo {pages}
  {1330001} (\bibinfo {year} {2013})}\BibitemShut {NoStop}%
\bibitem [{\citenamefont {Zhu}\ and\ \citenamefont {Wang}(2012)}]{zhu12}%
  \BibitemOpen
  \bibfield  {author} {\bibinfo {author} {\bibfnamefont {L.}~\bibnamefont
  {Zhu}}\ and\ \bibinfo {author} {\bibfnamefont {Q.}~\bibnamefont {Wang}},\
  }\href@noop {} {\bibfield  {journal} {\bibinfo  {journal} {Macromolecules}\
  }\textbf {\bibinfo {volume} {45}},\ \bibinfo {pages} {2937} (\bibinfo {year}
  {2012})}\BibitemShut {NoStop}%
\bibitem [{\citenamefont {Bellaiche}\ and\ \citenamefont
  {{\'I\~niguez}}(2013)}]{bellaiche13}%
  \BibitemOpen
  \bibfield  {author} {\bibinfo {author} {\bibfnamefont {L.}~\bibnamefont
  {Bellaiche}}\ and\ \bibinfo {author} {\bibfnamefont {J.}~\bibnamefont
  {{\'I\~niguez}}},\ }\href@noop {} {\bibfield  {journal} {\bibinfo  {journal}
  {Physical Review B}\ }\textbf {\bibinfo {volume} {88}},\ \bibinfo {pages}
  {014104} (\bibinfo {year} {2013})}\BibitemShut {NoStop}%
\bibitem [{\citenamefont {Sawaguchi}\ \emph {et~al.}(1951)\citenamefont
  {Sawaguchi}, \citenamefont {Maniwa},\ and\ \citenamefont
  {Hoshino}}]{sawaguchi51}%
  \BibitemOpen
  \bibfield  {author} {\bibinfo {author} {\bibfnamefont {E.}~\bibnamefont
  {Sawaguchi}}, \bibinfo {author} {\bibfnamefont {H.}~\bibnamefont {Maniwa}}, \
  and\ \bibinfo {author} {\bibfnamefont {S.}~\bibnamefont {Hoshino}},\ }\href
  {\doibase 10.1103/PhysRev.83.1078} {\bibfield  {journal} {\bibinfo  {journal}
  {Physical Review}\ }\textbf {\bibinfo {volume} {83}},\ \bibinfo {pages}
  {1078} (\bibinfo {year} {1951})}\BibitemShut {NoStop}%
\bibitem [{\citenamefont {Shirane}\ \emph {et~al.}(1951)\citenamefont
  {Shirane}, \citenamefont {Sawaguchi},\ and\ \citenamefont
  {Takagi}}]{shirane51}%
  \BibitemOpen
  \bibfield  {author} {\bibinfo {author} {\bibfnamefont {G.}~\bibnamefont
  {Shirane}}, \bibinfo {author} {\bibfnamefont {E.}~\bibnamefont {Sawaguchi}},
  \ and\ \bibinfo {author} {\bibfnamefont {Y.}~\bibnamefont {Takagi}},\ }\href
  {\doibase 10.1103/PhysRev.84.476} {\bibfield  {journal} {\bibinfo  {journal}
  {Physical Review}\ }\textbf {\bibinfo {volume} {84}},\ \bibinfo {pages} {476}
  (\bibinfo {year} {1951})}\BibitemShut {NoStop}%
\bibitem [{\citenamefont {Haun}\ \emph {et~al.}(1989)\citenamefont {Haun},
  \citenamefont {Harvin}, \citenamefont {Lanagan}, \citenamefont {Zhuang},
  \citenamefont {Jang},\ and\ \citenamefont {Cross}}]{haun89}%
  \BibitemOpen
  \bibfield  {author} {\bibinfo {author} {\bibfnamefont {M.~J.}\ \bibnamefont
  {Haun}}, \bibinfo {author} {\bibfnamefont {T.~J.}\ \bibnamefont {Harvin}},
  \bibinfo {author} {\bibfnamefont {M.~T.}\ \bibnamefont {Lanagan}}, \bibinfo
  {author} {\bibfnamefont {Z.~Q.}\ \bibnamefont {Zhuang}}, \bibinfo {author}
  {\bibfnamefont {S.~J.}\ \bibnamefont {Jang}}, \ and\ \bibinfo {author}
  {\bibfnamefont {L.~E.}\ \bibnamefont {Cross}},\ }\href {\doibase
  http://dx.doi.org/10.1063/1.342668} {\bibfield  {journal} {\bibinfo
  {journal} {Journal of Applied Physics}\ }\textbf {\bibinfo {volume} {65}},\
  \bibinfo {pages} {3173} (\bibinfo {year} {1989})}\BibitemShut {NoStop}%
\bibitem [{\citenamefont {Liu}\ and\ \citenamefont {Dkhil}(2011)}]{liu11}%
  \BibitemOpen
  \bibfield  {author} {\bibinfo {author} {\bibfnamefont {H.}~\bibnamefont
  {Liu}}\ and\ \bibinfo {author} {\bibfnamefont {B.}~\bibnamefont {Dkhil}},\
  }\href@noop {} {\bibfield  {journal} {\bibinfo  {journal} {Zeitschrift
  f{\"u}r Kristallographie}\ }\textbf {\bibinfo {volume} {226}},\ \bibinfo
  {pages} {163} (\bibinfo {year} {2011})}\BibitemShut {NoStop}%
\bibitem [{\citenamefont {Tagantsev}\ \emph {et~al.}(2013)\citenamefont
  {Tagantsev}, \citenamefont {Vaideeswaran}, \citenamefont {Vakhrushev},
  \citenamefont {Filimonov}, \citenamefont {Shaganov}, \citenamefont
  {Andronikova}, \citenamefont {Rudskoy}, \citenamefont {Baron}, \citenamefont
  {Uchiyama}, \citenamefont {Chernyshov}, \citenamefont {Bosak}, \citenamefont
  {Ujma}, \citenamefont {Roleder}, \citenamefont {Majchrowski}, \citenamefont
  {Ko}, \citenamefont {Setter},\ and\ \citenamefont {Burkovsky}}]{tagantsev13}%
  \BibitemOpen
  \bibfield  {author} {\bibinfo {author} {\bibfnamefont {A.~K.}\ \bibnamefont
  {Tagantsev}}, \bibinfo {author} {\bibfnamefont {K.}~\bibnamefont
  {Vaideeswaran}}, \bibinfo {author} {\bibfnamefont {S.~B.}\ \bibnamefont
  {Vakhrushev}}, \bibinfo {author} {\bibfnamefont {A.~V.}\ \bibnamefont
  {Filimonov}}, \bibinfo {author} {\bibfnamefont {A.}~\bibnamefont {Shaganov}},
  \bibinfo {author} {\bibfnamefont {D.}~\bibnamefont {Andronikova}}, \bibinfo
  {author} {\bibfnamefont {A.~I.}\ \bibnamefont {Rudskoy}}, \bibinfo {author}
  {\bibfnamefont {A.~Q.~R.}\ \bibnamefont {Baron}}, \bibinfo {author}
  {\bibfnamefont {H.}~\bibnamefont {Uchiyama}}, \bibinfo {author}
  {\bibfnamefont {D.}~\bibnamefont {Chernyshov}}, \bibinfo {author}
  {\bibfnamefont {A.}~\bibnamefont {Bosak}}, \bibinfo {author} {\bibfnamefont
  {Z.}~\bibnamefont {Ujma}}, \bibinfo {author} {\bibfnamefont {K.}~\bibnamefont
  {Roleder}}, \bibinfo {author} {\bibfnamefont {A.}~\bibnamefont
  {Majchrowski}}, \bibinfo {author} {\bibfnamefont {J.~H.}\ \bibnamefont {Ko}},
  \bibinfo {author} {\bibfnamefont {N.}~\bibnamefont {Setter}}, \ and\ \bibinfo
  {author} {\bibfnamefont {R.~G.}\ \bibnamefont {Burkovsky}},\ }\href@noop {}
  {\bibfield  {journal} {\bibinfo  {journal} {Nature Communications}\ }\textbf
  {\bibinfo {volume} {4}},\ \bibinfo {pages} {2229} (\bibinfo {year}
  {2013})}\BibitemShut {NoStop}%
\bibitem [{\citenamefont {Hlinka}\ \emph {et~al.}(2014)\citenamefont {Hlinka},
  \citenamefont {Ostapchuk}, \citenamefont {Buixaderas}, \citenamefont
  {Kadlec}, \citenamefont {Kuzel}, \citenamefont {Gregora}, \citenamefont
  {Kroupa}, \citenamefont {Savinov}, \citenamefont {Klic}, \citenamefont
  {Drahokoupil}, \citenamefont {Etxebarria},\ and\ \citenamefont
  {Dec}}]{hlinka14}%
  \BibitemOpen
  \bibfield  {author} {\bibinfo {author} {\bibfnamefont {J.}~\bibnamefont
  {Hlinka}}, \bibinfo {author} {\bibfnamefont {T.}~\bibnamefont {Ostapchuk}},
  \bibinfo {author} {\bibfnamefont {E.}~\bibnamefont {Buixaderas}}, \bibinfo
  {author} {\bibfnamefont {C.}~\bibnamefont {Kadlec}}, \bibinfo {author}
  {\bibfnamefont {P.}~\bibnamefont {Kuzel}}, \bibinfo {author} {\bibfnamefont
  {I.}~\bibnamefont {Gregora}}, \bibinfo {author} {\bibfnamefont
  {J.}~\bibnamefont {Kroupa}}, \bibinfo {author} {\bibfnamefont
  {M.}~\bibnamefont {Savinov}}, \bibinfo {author} {\bibfnamefont
  {A.}~\bibnamefont {Klic}}, \bibinfo {author} {\bibfnamefont {J.}~\bibnamefont
  {Drahokoupil}}, \bibinfo {author} {\bibfnamefont {I.}~\bibnamefont
  {Etxebarria}}, \ and\ \bibinfo {author} {\bibfnamefont {J.}~\bibnamefont
  {Dec}},\ }\href@noop {} {\bibfield  {journal} {\bibinfo  {journal} {Physical
  Review Letters}\ }\textbf {\bibinfo {volume} {112}},\ \bibinfo {pages}
  {197601} (\bibinfo {year} {2014})}\BibitemShut {NoStop}%
\bibitem [{\citenamefont {Teslic}\ and\ \citenamefont
  {Egami}(1998)}]{teslic98}%
  \BibitemOpen
  \bibfield  {author} {\bibinfo {author} {\bibfnamefont {S.}~\bibnamefont
  {Teslic}}\ and\ \bibinfo {author} {\bibfnamefont {T.}~\bibnamefont {Egami}},\
  }\href {\doibase 10.1107/S0108768198003802} {\bibfield  {journal} {\bibinfo
  {journal} {Acta Crystallographica B}\ }\textbf {\bibinfo {volume} {54}},\
  \bibinfo {pages} {750} (\bibinfo {year} {1998})}\BibitemShut {NoStop}%
\bibitem [{\citenamefont {Fujishita}\ \emph {et~al.}(2003)\citenamefont
  {Fujishita}, \citenamefont {Ishikawa}, \citenamefont {Tanaka}, \citenamefont
  {Ogawaguchi},\ and\ \citenamefont {Katano}}]{fujishita03}%
  \BibitemOpen
  \bibfield  {author} {\bibinfo {author} {\bibfnamefont {H.}~\bibnamefont
  {Fujishita}}, \bibinfo {author} {\bibfnamefont {Y.}~\bibnamefont {Ishikawa}},
  \bibinfo {author} {\bibfnamefont {S.}~\bibnamefont {Tanaka}}, \bibinfo
  {author} {\bibfnamefont {A.}~\bibnamefont {Ogawaguchi}}, \ and\ \bibinfo
  {author} {\bibfnamefont {S.}~\bibnamefont {Katano}},\ }\href {\doibase
  10.1143/JPSJ.72.1426} {\bibfield  {journal} {\bibinfo  {journal} {Journal of
  the Physical Society of Japan}\ }\textbf {\bibinfo {volume} {72}},\ \bibinfo
  {pages} {1426} (\bibinfo {year} {2003})}\BibitemShut {NoStop}%
\bibitem [{sup()}]{supp}%
  \BibitemOpen
  \href@noop {} {}\bibinfo {note} {See Supplemental Material, which includes
  Refs.
  \cite{kresse96,kresse99,perdew08,blochl94,abinit09,gonze97,troullier91,fhi98PP,piskunov07,fujishita03,ghosez99,tagantsev13},
  for a description of the simulation methods employed and complementary
  results and commentary.}\BibitemShut {Stop}%
\bibitem [{\citenamefont {Singh}(1995)}]{singh95}%
  \BibitemOpen
  \bibfield  {author} {\bibinfo {author} {\bibfnamefont {D.~J.}\ \bibnamefont
  {Singh}},\ }\href {\doibase 10.1103/PhysRevB.52.12559} {\bibfield  {journal}
  {\bibinfo  {journal} {Physical Review B}\ }\textbf {\bibinfo {volume} {52}},\
  \bibinfo {pages} {12559} (\bibinfo {year} {1995})}\BibitemShut {NoStop}%
\bibitem [{\citenamefont {Piskunov}\ \emph {et~al.}(2007)\citenamefont
  {Piskunov}, \citenamefont {Gopeyenko}, \citenamefont {Kotomin}, \citenamefont
  {Zhukovskii},\ and\ \citenamefont {Ellis}}]{piskunov07}%
  \BibitemOpen
  \bibfield  {author} {\bibinfo {author} {\bibfnamefont {S.}~\bibnamefont
  {Piskunov}}, \bibinfo {author} {\bibfnamefont {A.}~\bibnamefont {Gopeyenko}},
  \bibinfo {author} {\bibfnamefont {E.}~\bibnamefont {Kotomin}}, \bibinfo
  {author} {\bibfnamefont {Y.}~\bibnamefont {Zhukovskii}}, \ and\ \bibinfo
  {author} {\bibfnamefont {D.}~\bibnamefont {Ellis}},\ }\href {\doibase
  10.1016/j.commatsci.2007.03.012} {\bibfield  {journal} {\bibinfo  {journal}
  {Computational Materials Science}\ }\textbf {\bibinfo {volume} {41}},\
  \bibinfo {pages} {195} (\bibinfo {year} {2007})}\BibitemShut {NoStop}%
\bibitem [{iso()}]{isotropy}%
  \BibitemOpen
  \href@noop {} {\enquote {\bibinfo {title} {Isotropy software suite},}\
  }\bibinfo {howpublished} {\url{iso.byu.edu}}\BibitemShut {NoStop}%
\bibitem [{fnl()}]{fnlabels}%
  \BibitemOpen
  \href@noop {} {}\bibinfo {note} {The symmetry labels correspond to the
  setting with the origin at the {\sl B} site.}\BibitemShut {Stop}%
\bibitem [{\citenamefont {Prosandeev}\ \emph {et~al.}(2014)\citenamefont
  {Prosandeev}, \citenamefont {Xu}, \citenamefont {Faye}, \citenamefont {Duan},
  \citenamefont {Liu}, \citenamefont {Dkhil}, \citenamefont {Janolin},
  \citenamefont {{\'I\~niguez}},\ and\ \citenamefont
  {Bellaiche}}]{prosandeev14}%
  \BibitemOpen
  \bibfield  {author} {\bibinfo {author} {\bibfnamefont {S.}~\bibnamefont
  {Prosandeev}}, \bibinfo {author} {\bibfnamefont {C.}~\bibnamefont {Xu}},
  \bibinfo {author} {\bibfnamefont {R.}~\bibnamefont {Faye}}, \bibinfo {author}
  {\bibfnamefont {W.}~\bibnamefont {Duan}}, \bibinfo {author} {\bibfnamefont
  {H.}~\bibnamefont {Liu}}, \bibinfo {author} {\bibfnamefont {B.}~\bibnamefont
  {Dkhil}}, \bibinfo {author} {\bibfnamefont {P.~E.}\ \bibnamefont {Janolin}},
  \bibinfo {author} {\bibfnamefont {J.}~\bibnamefont {{\'I\~niguez}}}, \ and\
  \bibinfo {author} {\bibfnamefont {L.}~\bibnamefont {Bellaiche}},\ }\href@noop
  {} {\bibfield  {journal} {\bibinfo  {journal} {Physical Review B}\ }\textbf
  {\bibinfo {volume} {89}},\ \bibinfo {pages} {214111} (\bibinfo {year}
  {2014})}\BibitemShut {NoStop}%
\bibitem [{fn4()}]{fn4thorder}%
  \BibitemOpen
  \href@noop {} {}\bibinfo {note} {It may seem surprising that a 4th-order
  model can reproduce so well the energetics of a material that undergoes a
  first-order transition. However, this situation is typical among perovskite
  oxides; see for example \cite{iniguez01}.}\BibitemShut {Stop}%
\bibitem [{\citenamefont {Shirane}(1952)}]{shirane52}%
  \BibitemOpen
  \bibfield  {author} {\bibinfo {author} {\bibfnamefont {G.}~\bibnamefont
  {Shirane}},\ }\href {\doibase 10.1103/PhysRev.86.219} {\bibfield  {journal}
  {\bibinfo  {journal} {Physical Review}\ }\textbf {\bibinfo {volume} {86}},\
  \bibinfo {pages} {219} (\bibinfo {year} {1952})}\BibitemShut {NoStop}%
\bibitem [{\citenamefont {Kagimura}\ and\ \citenamefont
  {Singh}(2008)}]{kagimura08}%
  \BibitemOpen
  \bibfield  {author} {\bibinfo {author} {\bibfnamefont {R.}~\bibnamefont
  {Kagimura}}\ and\ \bibinfo {author} {\bibfnamefont {D.~J.}\ \bibnamefont
  {Singh}},\ }\href@noop {} {\bibfield  {journal} {\bibinfo  {journal}
  {Physical Review B}\ }\textbf {\bibinfo {volume} {77}},\ \bibinfo {pages}
  {104113} (\bibinfo {year} {2008})}\BibitemShut {NoStop}%
\bibitem [{\citenamefont {Reyes-Lillo}\ and\ \citenamefont
  {Rabe}(2013)}]{reyeslillo13}%
  \BibitemOpen
  \bibfield  {author} {\bibinfo {author} {\bibfnamefont {S.~E.}\ \bibnamefont
  {Reyes-Lillo}}\ and\ \bibinfo {author} {\bibfnamefont {K.~M.}\ \bibnamefont
  {Rabe}},\ }\href@noop {} {\bibfield  {journal} {\bibinfo  {journal} {Physical
  Review B}\ }\textbf {\bibinfo {volume} {88}},\ \bibinfo {pages} {180102}
  (\bibinfo {year} {2013})}\BibitemShut {NoStop}%
\bibitem [{\citenamefont {Ghosez}\ \emph {et~al.}(1999)\citenamefont {Ghosez},
  \citenamefont {Cockayne}, \citenamefont {Waghmare},\ and\ \citenamefont
  {Rabe}}]{ghosez99}%
  \BibitemOpen
  \bibfield  {author} {\bibinfo {author} {\bibfnamefont {P.}~\bibnamefont
  {Ghosez}}, \bibinfo {author} {\bibfnamefont {E.}~\bibnamefont {Cockayne}},
  \bibinfo {author} {\bibfnamefont {U.~V.}\ \bibnamefont {Waghmare}}, \ and\
  \bibinfo {author} {\bibfnamefont {K.~M.}\ \bibnamefont {Rabe}},\ }\href@noop
  {} {\bibfield  {journal} {\bibinfo  {journal} {Physical Review B}\ }\textbf
  {\bibinfo {volume} {60}},\ \bibinfo {pages} {836} (\bibinfo {year}
  {1999})}\BibitemShut {NoStop}%
\bibitem [{fnP()}]{fnPnma}%
  \BibitemOpen
  \href@noop {} {}\bibinfo {note} {The preference for the $Pbam$ ground state
  over the $Pnma$ ($a^{-}a^{-}c^{+}$) phase is not obvious from
  Fig.~\ref{fig3}(b). Hence, it seems to rely on further anharmonic couplings
  (inter-mode and involving strain) not reflected in the shown phonon bands.
  See \cite{supp} for further comments on this.}\BibitemShut {Stop}%
\bibitem [{\citenamefont {Catalan}\ and\ \citenamefont
  {Scott}(2009)}]{catalan09}%
  \BibitemOpen
  \bibfield  {author} {\bibinfo {author} {\bibfnamefont {G.}~\bibnamefont
  {Catalan}}\ and\ \bibinfo {author} {\bibfnamefont {J.~F.}\ \bibnamefont
  {Scott}},\ }\href {\doibase 10.1002/adma.200802849} {\bibfield  {journal}
  {\bibinfo  {journal} {Advanced Materials}\ }\textbf {\bibinfo {volume}
  {21}},\ \bibinfo {pages} {2463} (\bibinfo {year} {2009})}\BibitemShut
  {NoStop}%
\bibitem [{\citenamefont {{Di\'eguez}}\ \emph {et~al.}(2011)\citenamefont
  {{Di\'eguez}}, \citenamefont {{Gonz\'alez-V\'azquez}}, \citenamefont
  {Wojde{\l}},\ and\ \citenamefont {{\'I\~niguez}}}]{dieguez11}%
  \BibitemOpen
  \bibfield  {author} {\bibinfo {author} {\bibfnamefont {O.}~\bibnamefont
  {{Di\'eguez}}}, \bibinfo {author} {\bibfnamefont {O.~E.}\ \bibnamefont
  {{Gonz\'alez-V\'azquez}}}, \bibinfo {author} {\bibfnamefont {J.~C.}\
  \bibnamefont {Wojde{\l}}}, \ and\ \bibinfo {author} {\bibfnamefont
  {J.}~\bibnamefont {{\'I\~niguez}}},\ }\href@noop {} {\bibfield  {journal}
  {\bibinfo  {journal} {Physical Review B}\ }\textbf {\bibinfo {volume} {83}},\
  \bibinfo {pages} {094105} (\bibinfo {year} {2011})}\BibitemShut {NoStop}%
\bibitem [{\citenamefont {Prosandeev}\ \emph {et~al.}(2013)\citenamefont
  {Prosandeev}, \citenamefont {Wang}, \citenamefont {Ren}, \citenamefont
  {{\'I\~niguez}},\ and\ \citenamefont {Bellaiche}}]{prosandeev13}%
  \BibitemOpen
  \bibfield  {author} {\bibinfo {author} {\bibfnamefont {S.}~\bibnamefont
  {Prosandeev}}, \bibinfo {author} {\bibfnamefont {D.}~\bibnamefont {Wang}},
  \bibinfo {author} {\bibfnamefont {W.}~\bibnamefont {Ren}}, \bibinfo {author}
  {\bibfnamefont {J.}~\bibnamefont {{\'I\~niguez}}}, \ and\ \bibinfo {author}
  {\bibfnamefont {L.}~\bibnamefont {Bellaiche}},\ }\href@noop {} {\bibfield
  {journal} {\bibinfo  {journal} {Advanced Functional Materials}\ }\textbf
  {\bibinfo {volume} {23}},\ \bibinfo {pages} {234} (\bibinfo {year}
  {2013})}\BibitemShut {NoStop}%
\bibitem [{\citenamefont {Cazorla}\ and\ \citenamefont
  {{\'I\~niguez}}(2013)}]{cazorla13}%
  \BibitemOpen
  \bibfield  {author} {\bibinfo {author} {\bibfnamefont {C.}~\bibnamefont
  {Cazorla}}\ and\ \bibinfo {author} {\bibfnamefont {J.}~\bibnamefont
  {{\'I\~niguez}}},\ }\href@noop {} {\bibfield  {journal} {\bibinfo  {journal}
  {Physical Review B}\ }\textbf {\bibinfo {volume} {88}},\ \bibinfo {pages}
  {214430} (\bibinfo {year} {2013})}\BibitemShut {NoStop}%
\bibitem [{\citenamefont {Momma}\ and\ \citenamefont {Izumi}(2008)}]{vesta}%
  \BibitemOpen
  \bibfield  {author} {\bibinfo {author} {\bibfnamefont {K.}~\bibnamefont
  {Momma}}\ and\ \bibinfo {author} {\bibfnamefont {F.}~\bibnamefont {Izumi}},\
  }\href@noop {} {\bibfield  {journal} {\bibinfo  {journal} {Journal of Applied
  Crystallography}\ }\textbf {\bibinfo {volume} {41}},\ \bibinfo {pages} {653}
  (\bibinfo {year} {2008})}\BibitemShut {NoStop}%
\bibitem [{\citenamefont {Kresse}\ and\ \citenamefont
  {{Furthm\"uller}}(1996)}]{kresse96}%
  \BibitemOpen
  \bibfield  {author} {\bibinfo {author} {\bibfnamefont {G.}~\bibnamefont
  {Kresse}}\ and\ \bibinfo {author} {\bibfnamefont {J.}~\bibnamefont
  {{Furthm\"uller}}},\ }\href@noop {} {\bibfield  {journal} {\bibinfo
  {journal} {Physical Review B}\ }\textbf {\bibinfo {volume} {54}},\ \bibinfo
  {pages} {11169} (\bibinfo {year} {1996})}\BibitemShut {NoStop}%
\bibitem [{\citenamefont {Kresse}\ and\ \citenamefont
  {Joubert}(1999)}]{kresse99}%
  \BibitemOpen
  \bibfield  {author} {\bibinfo {author} {\bibfnamefont {G.}~\bibnamefont
  {Kresse}}\ and\ \bibinfo {author} {\bibfnamefont {D.}~\bibnamefont
  {Joubert}},\ }\href@noop {} {\bibfield  {journal} {\bibinfo  {journal}
  {Physical Review B}\ }\textbf {\bibinfo {volume} {59}},\ \bibinfo {pages}
  {1758} (\bibinfo {year} {1999})}\BibitemShut {NoStop}%
\bibitem [{\citenamefont {Perdew}\ \emph {et~al.}(2008)\citenamefont {Perdew},
  \citenamefont {Ruzsinszky}, \citenamefont {Csonka}, \citenamefont {Vydrov},
  \citenamefont {Scuseria}, \citenamefont {Constantin}, \citenamefont {Zhou},\
  and\ \citenamefont {Burke}}]{perdew08}%
  \BibitemOpen
  \bibfield  {author} {\bibinfo {author} {\bibfnamefont {J.~P.}\ \bibnamefont
  {Perdew}}, \bibinfo {author} {\bibfnamefont {A.}~\bibnamefont {Ruzsinszky}},
  \bibinfo {author} {\bibfnamefont {G.~I.}\ \bibnamefont {Csonka}}, \bibinfo
  {author} {\bibfnamefont {O.~A.}\ \bibnamefont {Vydrov}}, \bibinfo {author}
  {\bibfnamefont {G.~E.}\ \bibnamefont {Scuseria}}, \bibinfo {author}
  {\bibfnamefont {L.~A.}\ \bibnamefont {Constantin}}, \bibinfo {author}
  {\bibfnamefont {X.}~\bibnamefont {Zhou}}, \ and\ \bibinfo {author}
  {\bibfnamefont {K.}~\bibnamefont {Burke}},\ }\href {\doibase
  10.1103/PhysRevLett.100.136406} {\bibfield  {journal} {\bibinfo  {journal}
  {Physical Review Letters}\ }\textbf {\bibinfo {volume} {100}},\ \bibinfo
  {pages} {136406} (\bibinfo {year} {2008})}\BibitemShut {NoStop}%
\bibitem [{\citenamefont {Bl{\"o}chl}(1994)}]{blochl94}%
  \BibitemOpen
  \bibfield  {author} {\bibinfo {author} {\bibfnamefont {P.~E.}\ \bibnamefont
  {Bl{\"o}chl}},\ }\href@noop {} {\bibfield  {journal} {\bibinfo  {journal}
  {Physical Review B}\ }\textbf {\bibinfo {volume} {50}},\ \bibinfo {pages}
  {17953} (\bibinfo {year} {1994})}\BibitemShut {NoStop}%
\bibitem [{\citenamefont {Gonze}\ \emph {et~al.}(2009)\citenamefont {Gonze},
  \citenamefont {Amadon}, \citenamefont {Anglade}, \citenamefont {Beuken},
  \citenamefont {Bottin}, \citenamefont {Boulanger}, \citenamefont {Bruneval},
  \citenamefont {Caliste}, \citenamefont {Caracas}, \citenamefont {Cote} \emph
  {et~al.}}]{abinit09}%
  \BibitemOpen
  \bibfield  {author} {\bibinfo {author} {\bibfnamefont {X.}~\bibnamefont
  {Gonze}}, \bibinfo {author} {\bibfnamefont {B.}~\bibnamefont {Amadon}},
  \bibinfo {author} {\bibfnamefont {P.~M.}\ \bibnamefont {Anglade}}, \bibinfo
  {author} {\bibfnamefont {J.~M.}\ \bibnamefont {Beuken}}, \bibinfo {author}
  {\bibfnamefont {F.}~\bibnamefont {Bottin}}, \bibinfo {author} {\bibfnamefont
  {P.}~\bibnamefont {Boulanger}}, \bibinfo {author} {\bibfnamefont
  {F.}~\bibnamefont {Bruneval}}, \bibinfo {author} {\bibfnamefont
  {D.}~\bibnamefont {Caliste}}, \bibinfo {author} {\bibfnamefont
  {R.}~\bibnamefont {Caracas}}, \bibinfo {author} {\bibfnamefont
  {M.}~\bibnamefont {Cote}},  \emph {et~al.},\ }\href@noop {} {\bibfield
  {journal} {\bibinfo  {journal} {Computer Physics Communications}\ }\textbf
  {\bibinfo {volume} {180}},\ \bibinfo {pages} {2582–2615} (\bibinfo {year}
  {2009})}\BibitemShut {NoStop}%
\bibitem [{\citenamefont {Gonze}\ and\ \citenamefont {Lee}(1997)}]{gonze97}%
  \BibitemOpen
  \bibfield  {author} {\bibinfo {author} {\bibfnamefont {X.}~\bibnamefont
  {Gonze}}\ and\ \bibinfo {author} {\bibfnamefont {C.}~\bibnamefont {Lee}},\
  }\href@noop {} {\bibfield  {journal} {\bibinfo  {journal} {Physical Review
  B}\ }\textbf {\bibinfo {volume} {55}},\ \bibinfo {pages} {10355} (\bibinfo
  {year} {1997})}\BibitemShut {NoStop}%
\bibitem [{\citenamefont {Troullier}\ and\ \citenamefont
  {Martins}(1991)}]{troullier91}%
  \BibitemOpen
  \bibfield  {author} {\bibinfo {author} {\bibfnamefont {N.}~\bibnamefont
  {Troullier}}\ and\ \bibinfo {author} {\bibfnamefont {J.~L.}\ \bibnamefont
  {Martins}},\ }\href@noop {} {\bibfield  {journal} {\bibinfo  {journal}
  {Physical Review B}\ }\textbf {\bibinfo {volume} {43}},\ \bibinfo {pages}
  {1993} (\bibinfo {year} {1991})}\BibitemShut {NoStop}%
\bibitem [{\citenamefont {Fuchs}\ and\ \citenamefont
  {Scheffler}(1999)}]{fhi98PP}%
  \BibitemOpen
  \bibfield  {author} {\bibinfo {author} {\bibfnamefont {M.}~\bibnamefont
  {Fuchs}}\ and\ \bibinfo {author} {\bibfnamefont {M.}~\bibnamefont
  {Scheffler}},\ }\href@noop {} {\bibfield  {journal} {\bibinfo  {journal}
  {Computer Physics Communications}\ }\textbf {\bibinfo {volume} {119}},\
  \bibinfo {pages} {67} (\bibinfo {year} {1999})}\BibitemShut {NoStop}%
\bibitem [{\citenamefont {{\'I\~niguez}}\ \emph {et~al.}(2001)\citenamefont
  {{\'I\~niguez}}, \citenamefont {Ivantchev}, \citenamefont {{P\'erez}-Mato},\
  and\ \citenamefont {{Garc\'{\i}a}}}]{iniguez01}%
  \BibitemOpen
  \bibfield  {author} {\bibinfo {author} {\bibfnamefont {J.}~\bibnamefont
  {{\'I\~niguez}}}, \bibinfo {author} {\bibfnamefont {S.}~\bibnamefont
  {Ivantchev}}, \bibinfo {author} {\bibfnamefont {J.~M.}\ \bibnamefont
  {{P\'erez}-Mato}}, \ and\ \bibinfo {author} {\bibfnamefont {A.}~\bibnamefont
  {{Garc\'{\i}a}}},\ }\href {\doibase 10.1103/PhysRevB.63.144103} {\bibfield
  {journal} {\bibinfo  {journal} {Physical Review B}\ }\textbf {\bibinfo
  {volume} {63}},\ \bibinfo {pages} {144103} (\bibinfo {year}
  {2001})}\BibitemShut {NoStop}%
\end{thebibliography}

%

\end{document}